\documentclass{article}

\pdfoutput=1
\usepackage[utf8]{inputenc}
\usepackage[backend=biber,
            bibstyle=phys,
            sorting=none,
            url=false,
            doi=false,
            eprint=true]{biblatex}
\DeclareFieldFormat*{title}{\mkbibitalic{#1}}
\DeclareFieldFormat{labelnumberwidth}{\mkbibbrackets{#1}~~}
\usepackage{amssymb,amsmath,latexsym}
\usepackage[]{hyperref}\hypersetup{}
\usepackage{euscript,graphics,xcolor,tikz}
\usetikzlibrary{arrows,positioning}

\numberwithin{equation}{section}

\renewcommand{\sl}{\mathfrak{sl}}

\newcommand\mI{\mathcal{I}}
\newcommand\mA{\mathcal{A}}
\newcommand\mD{\mathcal{D}}
\newtheorem{prop}{Proposition}[section]
\newtheorem{rmk}{Remark}[section]
\newcommand{\RED}[1]{{\color{black}#1}}

\catcode`,\active

\catcode`\,12

\begin{document}

\title{\bf Sklyanin-like algebras for ($q$-)linear grids and ($q$-)para-Krawtchouk
polynomials}
\author{
Geoffroy Bergeron\textsuperscript{$1$}
\footnote{E-mail: geoffroy.bergeron@umontreal.ca}~,
Julien Gaboriaud\textsuperscript{$1$}
\footnote{E-mail: julien.gaboriaud@umontreal.ca}~,
Luc Vinet\textsuperscript{$1$}
\footnote{E-mail: vinet@CRM.UMontreal.CA}~,
Alexei Zhedanov\textsuperscript{$2$}
\footnote{E-mail: zhedanov@yahoo.com}
\\[.5em]
\textsuperscript{$1$}
\small~Centre de Recherches Math\'ematiques,
Universit\'e de Montr\'eal, \\
\small~P.O. Box 6128, Centre-ville Station,
Montr\'eal (Qu\'ebec), H3C 3J7,
Canada.\\[.9em]
\textsuperscript{$2$}\small~School of Mathematics,
Renmin University of China,
Beijing, 100872, China.
}
\date{\today}
\maketitle

\hrule
\begin{abstract}

\noindent S-Heun operators on linear and $q$-linear grids are introduced. These operators
are special cases of Heun operators and are related to Sklyanin-like algebras. The
Continuous Hahn and Big $q$-Jacobi polynomials are functions on which these S-Heun
operators have natural actions. We show that the S-Heun operators encompass both the
bispectral operators and Kalnins and Miller's structure operators. These four structure
operators realize special limit cases of the trigonometric degeneration of the original
Sklyanin algebra. Finite-dimensional representations  of these algebras are obtained from
a truncation condition. The corresponding representation bases are finite families of
polynomials: the para-Krawtchouk and $q$-para-Krawtchouk ones. A natural algebraic
interpretation of these polynomials that had been missing is thus obtained. We also
recover the Heun operators attached to the corresponding bispectral problems as quadratic
combinations of the S-Heun operators.\\

\end{abstract}
\hrule


\section*{Introduction}\label{sec:intro}

In the study of orthogonal polynomials (OPs), many of their properties are expressed as
structure relations between family members with different parameters, arguments or
degrees, examples are the three term recurrence relation, the differential/difference
equation, the backward/forward relation, etc. As it turns out, the operators involved in
these formulas realize algebras that synthesize much of the characterization of these
polynomial ensembles.  The present paper relates to this framework.

One such instance that has proven very fruitful is the (algebraic) study of the two
bispectral operators associated to hypergeometric OPs. These operators are the recurrence
and the differential/difference operators.  Let us focus on the developments related to
the Askey--Wilson polynomials; since these polynomials sit at the top of the Askey scheme,
the gist of their description descends onto all the lower families in the scheme. The two
bispectral operators for the Askey--Wilson polynomials do not commute: they form an algebra
whose relations have been found by Zhedanov in \cite{Zhedanov1991} and it is
usually referred to as the Askey--Wilson algebra.

This algebra has appeared in a great variety of contexts, such as knot theory
\cite{BullockPrzytycki1999}, double affine Hecke algebras and representation theory
\cite{Koornwinder2007a, Koornwinder2008, Mazzocco2016}, Howe duality
\cite{FrappatGaboriaudetal2020, GaboriaudVinetetal2020}, integrable models
\cite{Baseilhac2005, Baseilhac2005a, BaseilhacKoizumi2005, VinetZhedanov2008}, algebraic
combinatorics \cite{TerwilligerVidunas2004, Terwilliger2011, Terwilliger2013,
Terwilliger2018}, the Racah problem for $U_q(\sl_2)$ \cite{GranovskiiZhedanov1993,
Huang2016}, etc. The abovementioned connections have some specializations for all entries
of the Askey tableau.

The work of Kalnins and Miller \cite{KalninsMiller1989, KalninsMiller1987, Miller1987}
based on the use of four structure or contiguity operators is another approach that
illustrates the use of symmetry techniques in the study of OPs. These operators that shall
be referred to as structure operators in the following
correspond to the backward and forward operators, as well as to another pair of operators
that ``factorize'' \cite{InfeldHull1951} the differential/difference operator.  It was
recently observed \cite{Koornwinder2007} that for the Askey--Wilson polynomials, these
operators realize the relations of the trigonometric degeneration
\cite{GorskyZabrodin1993} of the Sklyanin algebra \cite{Sklyanin1983}. To our knowledge,
the Sklyanin-like algebras similarly connected to other families of OPs have not been
described so far and will be the center of attention here.

The differential/difference operator of which the OPs are eigenfunctions belongs to the
intersection of the sets of operators involved in the two pictures.  A natural question is
the following: \textit{what is the most elementary set of operators that encompasses all
operators in both of the approaches above?} In the case of the Askey--Wilson polynomials,
this answer was given in \cite{GaboriaudTsujimotoetal2020}: it is the set of so-called
S-Heun operators on the Askey--Wilson grid (these are special types of Heun operators that
will be defined in the next section).  Operators of the Heun type are related to the
tridiagonalization procedure \cite{IsmailKoelink2012, GrunbaumVinetetal2017} and have been
given an algebraic formulation \cite{Grunbaum2001, GrunbaumVinetetal2018}. They have been
identified as Hamiltonians of quantum Euler--Arnold tops \cite{Turbiner2016}, they have
been connected to band-time limiting \cite{Slepian1983, Landau1985} and to the study of
entanglement in spin chains \cite{CrampeNepomechieetal2019, CrampeNepomechieetal2020} and
they have been studied quite a lot recently \cite{Takemura2017, Takemura2018,
BaseilhacTsujimotoetal2019, BaseilhacPimenta2019, VinetZhedanov2019, CrampeVinetetal2020,
TsujimotoVinetetal2019, BaseilhacVinetetal2020, BergeronVinetetal2020,
BergeronCrampeetal2020}.  As will be shown below, the S-Heun operators allow a
factorization of these Heun operators.  Let us note that in addition to the unification of
the two approaches described above, the S-Heun framework has also led to a novel algebraic
interpretation of the $q$-para-Racah polynomials.  The goal of the present paper is to
look at the grids of linear type from the S-Heun operators point of view.  As a byproduct,
an algebraic interpretation of the para-Krawtchouk and $q$-para-Krawtchouk polynomials
will be obtained. These polynomials were first identified in the context of perfect state
transfer and fractional revival on quantum spin chains \cite{VinetZhedanov2012,
LemayVinetetal2016, GenestVinetetal2016a, BosseVinet2017} and their algebraic
interpretation was still lacking.

We will introduce the S-Heun operators on linear grids in Section \ref{sec:lintype}. The
simplest example of operators of this type will be worked out in Section \ref{sec:contlin}
(this will involve differential operators, the Jacobi polynomials and the ordinary Heun
operator).  Section \ref{sec:sheun} will focus on the S-Heun operators on the discrete
linear grid. A new degeneration of the Sklyanin algebra will be presented.  Of relevance
in this case, the Continuous Hahn polynomials will be seen to truncate to the
para-Krawtchouk polynomials under a special condition and an algebraic interpretation of
such a truncation will be given. The Heun operator on the uniform grid will also be
recovered.  The $q$-linear grid will be examined in Section \ref{sec:qlin} and the
previous analysis will be repeated. The degeneration of the Sklyanin algebra that arises
will be identified as $U_q(\sl_2)$. The Big $q$-Jacobi polynomials will be involved, and
they will be observed to reduce to the $q$-para-Krawtchouk polynomials under a certain
condition. The Big $q$-Jacobi Heun operator will also be recovered as well.  Connections
between the three grids and the associated S-Heun operators and Sklyanin-type algebras
will be presented in Section \ref{sec:connection}, followed by concluding remarks. The
quadratic relations between the S-Heun operators for the three different types of grids
are listed in Appendix \ref{sec:appendix}.

\section{S-Heun operators on linear-type grids}\label{sec:lintype}

S-Heun operators are defined as the most general second order differential/difference
operators without diagonal term that obey a degree raising condition.  Like Heun
operators, they can be defined on different grids.  We now introduce the three linear
grids that we will use and obtain the S-Heun operators associated to each.

\subsection{The discrete linear grid}\label{ssec:discrete_lin}

Consider the operator $S$
\begin{align}\label{S_def}
 S=A_1 T_+ + A_2 T_-
\end{align}
where
\begin{align}\label{TPM}
 T_+ f(x) = f(x+1), \qquad T_- f(x) = f(x-1)
\end{align}
are shift operators, and $A_{1,2}$ are functions in the real variable $x$.  Impose
that $S$ maps polynomials of degree $n$ onto polynomials of degree no higher than $n+1$,
namely,
\begin{align}\label{SPP}
S P_n(x) = \tilde{P}_{n+1}(x)
\end{align}
for all $n=0,1,2,\dots$. This defines the S-Heun operators on the discrete linear grid.

It is sufficient to enforce this raising condition on monomials $x^{n}$; for $n=0$ and
$n=1$, it reads
\begin{subequations}\label{eq:S0S1}
\begin{align}
A_1 + A_2 &= a_{00} + a_{01} x,\\
A_1(x+1) + A_2(x-1) &= a_{10} +a_{11} x + a_{12} x^2,
\end{align}
\end{subequations}
for some arbitrary parameters $a_{ij}$. This can be rewritten as
\begin{subequations}\label{eq:A1pA2A1mA2}
\begin{align}
A_1 + A_2 &= a_{00} + a_{01} x,\\
A_1 - A_2 &= a_{10} + (a_{11}-a_{00}) x + (a_{12}-a_{01}) x^2.
\end{align}
\end{subequations}
Straightforward induction shows that in general one has
\begin{align}\label{eq:Saction}
 Sx^{n}=A_1(x+1)^{n}+A_2(x-1)^{n}=\sum_{k=0}^{n}\binom{n}{k}x^{k}[A_1+(-1)^{n-k}A_2],
\end{align}
which is a polynomial of degree $n+1$.  Thus, the functions $A_1$, $A_2$
\begin{subequations}\label{eq:A1A2}
\begin{align}
 A_1 &= \tfrac{1}{2}\left[(-a_{01}+a_{12}){x}^{2}
 + (-a_{00}+a_{01}+a_{11})x + (a_{00}+a_{10})\right],\\
 A_2 &= \tfrac{1}{2}\left[(+a_{01}-a_{12}){x}^{2}
 + (+a_{00}+a_{01}-a_{11})x + (a_{00}-a_{10})\right]
\end{align}
\end{subequations}
satisfy \eqref{eq:A1pA2A1mA2} and the operator \eqref{S_def} meets the
degree raising condition.
\begin{prop}
With the functions $A_1$, $A_2$ given by \eqref{eq:A1A2}, the operator $S$
in \eqref{S_def} is the most general S-Heun operator on the linear grid. $S$ depends on
$5$ free parameters
and spans a $5$-dimensional linear space.
The elements
\begin{subequations}\label{eq:LM1M2R1R2_def}
\begin{align}
\label{L_def}
L &= \tfrac{1}{2} \left[T_+-T_-\right],\\
\label{M1_def}
M_1 &= \tfrac{1}{2} \left[T_++T_-\right],\\
\label{M2_def}
M_2 &= \tfrac{1}{2} x\left[T_+-T_-\right],\\
\label{R1_def}
R_1 &=\tfrac{1}{2} x\left[(1-2x)T_++(1+2x)T_-\right],\\
\label{R2_def}
R_2 &= \tfrac{1}{2} x\left[T_++T_-\right].
\end{align}
\end{subequations}
form a basis for this space.
\end{prop}
Using \eqref{eq:Saction}, one sees that the operator $L$ is a lowering operator (it lowers
by one the degree of polynomials in $x$), the operators $M_1$, $M_2$ are stabilizing
operators (they do not change the degree) and the operators $R_1$, $R_2$ are raising
operators (they raise it by one).

\subsection{The $q$-linear grid}\label{ssec:q_lin}

Condider now the $q$-linear grid $z=q^{x}$ (or exponential grid). The S-Heun operators
$\hat{S}$ on that grid are of the form
\begin{align}\label{eq:Shat_def}
 \hat{S}=\hat{A}_1(z,q)\hat{T}_++\hat{A}_2(z,q)\hat{T}_-,
\end{align}
with
\begin{align}\label{}
 \hat{T}_\pm f(z)=f(q^{\pm1}z),
\end{align}
and are taken to map polynomials in $z$ onto polynomials of at most one degree higher:
$\hat{S}P_n(z)=\tilde{P}_{n+1}(z)$. Imposing this degree raising condition on the first
monomials $1$ and $z$ yields
\begin{subequations}\label{eq:condA1qpmA2q}
\begin{align}
 \hat{A}_1(z,q)+\hat{A}_2(z,q)&=a_{00}+a_{01}z,\\
 \hat{A}_1(z,q)q+\hat{A}_2(z,q)q^{-1}&=a_{10}z^{-1}+a_{11}+a_{12}z.
\end{align}
\end{subequations}
Straightforward induction shows that in general one has
\begin{align}\label{eq:hat_Saction}
 \hat{S}z^{n}=(\hat{A}_1q^{n}+\hat{A}_2q^{-n})z^{n}=z^{n}[\hat{A}_1q+\hat{A}_2q^{-1}]
 \frac{q^{n}-q^{-n}}{q-q^{-1}}
 -z^{n}[\hat{A}_1+\hat{A}_2]\frac{q^{n-1}-q^{1-n}}{q-q^{-1}},
\end{align}
which is a polynomial of degree $n+1$ in $z$.  Thus, an operator $\hat{S}$ with
$\hat{A}_1(z,q)$ and $\hat{A}_2(z,q)$ that satisfies
\eqref{eq:condA1qpmA2q} will obey the degree raising condition on
any monomial. We hence obtain:
\begin{align}\label{eq:hat_A1A2}
 \hat{A}_1(z,q)=\hat{A}_2(z,q^{-1})
 =\frac{1}{(q-q^{-1})z}\left[a_{10}+(a_{11}-a_{00}q^{-1})z
 +(a_{12}-a_{01}q^{-1})z^{2}\right].
\end{align}
\begin{prop}
With the functions $\hat{A}_1(z,q)$, $\hat{A}_2(z,q)$ given by \eqref{eq:hat_A1A2}, the
operator $\hat{S}$ in \eqref{eq:Shat_def} is the most general S-Heun operator on the
$q$-linear grid. $\hat{S}$ depends on $5$ free parameters
and spans a $5$-dimensional linear space.  The elements
\begin{subequations}\label{eq:hatLM1M2R1R2_def}
\begin{align}
 \hat{L}&=\frac{1}{(q-q^{-1})}z^{-1}(\hat{T}_+-\hat{T}_-),\\
 \hat{M}_1&=\frac{1}{(q-q^{-1})}(-q^{-1}\hat{T}_++q\hat{T}_-),\\
 \hat{M}_2&=\frac{1}{(q-q^{-1})}(\hat{T}_+-\hat{T}_-),\\
 \hat{R}_1&=\frac{1}{(q-q^{-1})}z(-q^{-1}\hat{T}_++q\hat{T}_-),\\
 \hat{R}_2&=\frac{1}{(q-q^{-1})}z(\hat{T}_+-\hat{T}_-).
\end{align}
\end{subequations}
can be chosen as a basis for this space.
\end{prop}
Looking at \eqref{eq:hat_Saction} and \eqref{eq:hat_A1A2}, one sees that the operator
$\hat{L}$ lowers the degrees, and that the $\hat{M}_i$'s and the $\hat{R}_i$'s are
respectively stabilizing and raising operators.

\subsection{The simplest case: differential S-Heun operators}\label{ssec:diff}

The definition of the S-Heun operators on the real line goes as follows.  Consider the
first-order differential operator
\begin{align}\label{}
 \bar{S}=\bar{A}_1(x)\frac{d}{dx}+\bar{A}_2(x)
\end{align}
and impose the raising condition $\bar{S}p_n(x)=\tilde{p}_{n+1}(x)$ which demands that
$\bar{S}$ sends polynomials into polynomials of one degree higher. The general solution is
given by
\begin{align}\label{}
 \bar{A}_1(x)=a_{10}+a_{11}x+a_{12}x^{2},\qquad \bar{A}_2(x)=a_{20}+a_{21}x.
\end{align}
This leads to the following set of five linearly independent S-Heun operators
\cite{Turbiner2016}
\begin{align}\label{eq:barLM1M2R1R2}
 \bar{L}=\frac{d}{dx},\qquad
 \bar{M}_1=1,\qquad
 \bar{M}_2=x\frac{d}{dx},\qquad
 \bar{R}_1=x,\qquad
 \bar{R}_2=x^{2}\frac{d}{dx},
\end{align}
which are once again labelled according to their property of lowering ($\bar{L}$),
stabilizing ($\bar{M}$) or raising ($\bar{R}$) the degree of polynomials in the variable
$x$.

These S-Heun operators can also be obtained as a $q\to1$ limit of the ones defined on the
$q$-linear grid. More precisely, writing $q=e^{\hbar}$ and letting $\hbar\to0$, one
obtains
\begin{align}\label{eq:q1limit_LM1M2R1R2}
 \lim_{q\to1}\hat{L}=\bar{L},\qquad
 \lim_{q\to1}\hat{M}_1=\bar{M}_1-\bar{M}_2,\qquad
 \lim_{q\to1}\hat{M}_2=\bar{M}_2,\qquad
 \lim_{q\to1}\hat{R}_1=\bar{R}_1-\bar{R}_2,\qquad
 \lim_{q\to1}\hat{R}_2=\bar{R}_2.
\end{align}
This connects with the definition of the continuous S-Heun operators.  These
S-Heun operators will also be related to the ordinary Heun operator introduced in the next
section.

\section{The continuous case}\label{sec:contlin}

The goal of this section is to revisit (mostly known) results with a point of view that
will be adopted in the following sections.  Here, we are interested in studying the OPs
and algebras related to the set of the five S-Heun operators defined in Section
\ref{ssec:diff}.

\subsection{The stabilizing subalgebra}\label{}

We first study the subset $\{\bar{L}, \bar{M}_1, \bar{M}_2\}$ of S-Heun operators that
stabilize the set of polynomials of a given degree. Let us denote by $\bar{Q}$ the most
general quadratic combination of these operators. Using the relations of Appendix
\ref{sec:appendix}, it is always possible to reduce $\bar{Q}$ to an expression of the form
\begin{align}\label{}
 \bar{Q}=\alpha_1 \bar{L}^{2}+\alpha_2\bar{L}\bar{M}_1+\alpha_3\bar{L}\bar{M}_2
        +\alpha_4\bar{M}_1{\!}^{2}+\alpha_5\bar{M}_1\bar{M}_2+\alpha_6\bar{M}_2{\!}^{2}.
\end{align}
Using the realizations \eqref{eq:barLM1M2R1R2}, the eigenvalue equation for the
second-order differential operator $\bar{Q}$ can be brought in the form
\begin{align}\label{}
\begin{aligned}
 \bar{\mD}P_n^{(\alpha,\beta)}(x)&=n(n+\alpha+\beta+1)P_n^{(\alpha,\beta)}(x),\\
 \bar{\mD}&=(x^{2}-1)\frac{d^{2}}{dx^{2}}+[(\alpha-\beta)+(\alpha+\beta+2)x]\frac{d}{dx},
\end{aligned}
\end{align}
which is recognized as the differential equation satisfied by the Jacobi polynomials
\cite{KoekoekLeskyetal2010}.

We have thus identified the family of OPs related to these (ordinary) S-Heun operators,
and as will be seen in the next subsection, certain combinations of these S-Heun
operators provide the structure relations of these polynomials.

\subsection{Jacobi polynomials and their structure relations}\label{}

Consider the forward and backward operators for the Jacobi polynomials
\begin{subequations}\label{eq:barttsmms_def}
\begin{align}\label{}
 \bar{\tau}=\bar{L},\qquad
 \bar{\tau}^{(\alpha,\beta)^{*}}
 =-\bar{L}+(\alpha-\beta)\bar{M}_1+(\alpha+\beta)\bar{R}_1+\bar{R}_2.
\end{align}
and the contiguity operators
\begin{align}\label{}
 \bar{\mu}^{(\alpha)}=-\bar{L}+\alpha\bar{M}_1+\bar{M}_2,\qquad
 \bar{\mu}^{(\beta)^{*}}=\bar{L}+\beta\bar{M}_1+\bar{M}_2.
\end{align}
\end{subequations}
These four operators act very simply on the Jacobi polynomials:
\begin{subequations}\label{eq:barttsmms_act}
\begin{align}\label{}
 \bar{\tau}P_n^{(\alpha,\beta)}(x)
 &=\tfrac{1}{2}(n+\alpha+\beta+1)P_{n-1}^{(\alpha+1,\beta+1)}(x),\\
 \bar{\tau}^{(\alpha,\beta)^{*}}P_n^{(\alpha,\beta)}(x)
 &=2(n+1)P_{n+1}^{(\alpha-1,\beta-1)}(x),\\
 \bar{\mu}^{(\alpha)}P_n^{(\alpha,\beta)}(x)
 &=(n+\alpha)P_{n}^{(\alpha-1,\beta+1)}(x),\\
 \bar{\mu}^{(\beta)^{*}}P_n^{(\alpha,\beta)}(x)
 &=(n+\beta)P_{n}^{(\alpha+1,\beta-1)}(x).
\end{align}
\end{subequations}
The operators $\bar{\mu}^{(\alpha)}$, $\bar{\mu}^{(\beta)^{*}}$, $\bar{\tau}$,
$\bar{\tau}^{(\alpha,\beta)^{*}}$ built from linear combinations of S-Heun operators are
of the type studied by Kalnins and Miller \cite{KalninsMiller1987}.

We have mentioned in the introduction that S-Heun operators encompass both the structure
operators of Kalnins and Miller and the bispectral operators.  Let us indicate how the
latter operators appear in this context. First, as mentioned above, the Jacobi
differential operator appears as a quadratic combination of the stabilizing generators. We
can actually provide a factorization of this operator either as a product of two
contiguous operators or as the product of the forward and backward operator:
\begin{align}\label{}
\begin{aligned}
 \bar{\mD}&=\bar{\mu}^{(\alpha+1)}\bar{\mu}^{(\beta)^{*}}-(\alpha+1)\beta\\
        &=\bar{\mu}^{(\beta+1)^{*}}\bar{\mu}^{(\alpha)}-\alpha(\beta+1)\\
        &=\bar{\tau}^{(\alpha+1,\beta+1)^{*}}\bar{\tau}\\
        &=\bar{\tau}\bar{\tau}^{(\alpha,\beta)^{*}}-(\alpha+\beta).
\end{aligned}
\end{align}
The other bispectral operator $\bar{X}$ is the multiplication by the variable $x$.
It can be directly expressed as $\bar{R}_1$, but since it will appear as a quadratic
combination of the S-Heun operators for other grids, we shall write it here as
\begin{align}\label{}
 \bar{X}=\bar{R}_1\bar{M}_1.
\end{align}
We have thus recovered the two bispectral operators as quadratic combinations in the
S-Heun operators. This completes the observation that the S-Heun operators are the
elementary blocks behind the two factorizations.

\subsection{The Sklyanin-like algebra realized by the structure operators}\label{}

We now focus on the algebras that are realized by these sets of operators.  On the one
hand the pair of bispectral Jacobi operators is known \cite{GenestIsmailetal2016} to
generate the Jacobi algebra that has been well studied \cite{GranovskiiLutzenkoetal1992}.
On the other hand, the algebra formed by the $4$ linear operators $\bar{\mu}^{(\alpha)}$,
$\bar{\mu}^{(\beta)^{*}}$, $\bar{\tau}$, $\bar{\tau}^{(\alpha,\beta)^{*}}$ can be seen to
be a degeneration of the Sklyanin algebra \cite{Sklyanin1983}.

We now give a presentation of this algebra. Denote
$\nu=-\tfrac{1}{2}(\alpha+\beta)$ and set
\begin{align}\label{eq:realsl2}
 \bar{A}=\bar{M}_2-\nu\bar{M}_1,\qquad
 \bar{B}=\bar{R}_2-2\nu\bar{R}_1,\qquad
 \bar{C}=\bar{L},\qquad
 \bar{D}=\bar{M}_1.
\end{align}
These linear combinations of $\bar{\mu}^{(\alpha)}$, $\bar{\mu}^{(\beta)^{*}}$,
$\bar{\tau}$, $\bar{\tau}^{(\alpha,\beta)^{*}}$ have been chosen in order to simplify
the relations.
\begin{prop}
The operators $\bar{A}$, $\bar{B}$, $\bar{C}$, $\bar{D}$ obey the homogeneous quadratic
relations
\begin{gather}\label{eq:barSkl4}
\begin{gathered}{}
 [\bar{C},\bar{D}]=0,\qquad
 [\bar{A},\bar{C}]=-\bar{C}\bar{D},\qquad
 [\bar{A},\bar{D}]=0,\\
 [\bar{B},\bar{C}]=-2\bar{A}\bar{D},\qquad
 [\bar{A},\bar{B}]=\bar{B}\bar{D},\qquad
 [\bar{B},\bar{D}]=0.
\end{gathered}
\end{gather}
\end{prop}
\begin{rmk}
One will notice that these relations are actually the relations of the $\sl_2$ Lie
algebra supplemented with a central element $D$
\RED{(one recovers $U(\sl_2)$ by quotienting the above algebra
\eqref{eq:barSkl4} by the additional relation $D=1$)}.
The reason why we wrote these in a quadratic fashion is to make easier the comparison with
the other Sklyanin algebras that will be obtained later.
\end{rmk}
One observes that if $\nu$ is an integer or half-integer, the realization
\eqref{eq:realsl2} is associated to a finite dimensional representation of dimension
$2\nu+1$.

%


%

\subsection{Recovering the Heun operator}\label{}

We now show how to recover the ordinary (differential) Heun operator from the knowledge of
the S-Heun operators.

The generic Heun operator $\bar{W}$ can be expressed as the most general
tridiagonalization of the hypergeometric operator \cite{GrunbaumVinetetal2017}. It has
been known to be
\begin{align}\label{eq:gen_contHeun}
 \bar{W}=Q_3(x)\frac{d^{2}}{dx^{2}}+Q_2(x)\frac{d}{dx}+Q_1(x),
\end{align}
where $Q_3(x)$, $Q_2(x)$ and $Q_1(x)$ are general polynomials of degree $3$, $2$ and $1$
respectively.

Let us consider the most general quadratic combination of S-Heun operators that does not
raise the degree of polynomials by more than one. Using the quadratic homogeneous
relations of Appendix \ref{sec:appendix}, it is always possible to simplify such an
expression to
\begin{align}\label{}
 \bar{W}=\alpha_1\bar{L}^{2}+\alpha_2\bar{L}\bar{M}_1+\alpha_3\bar{L}\bar{M}_2
        +\alpha_4\bar{M}_1{\!}^{2}+\alpha_5\bar{M}_1\bar{M}_2+\alpha_6\bar{M}_2{\!}^{2}
        +\beta_1\bar{M}_1\bar{R}_2+\beta_2\bar{M}_2\bar{R}_1+\beta_3\bar{M}_2\bar{R}_2.
\end{align}
From the differential expressions of the generators we obtain
\begin{align}\label{}
\begin{aligned}
 \bar{W}&=Q_3(x)\frac{d^{2}}{dx^{2}}+Q_2(x)\frac{d}{dx}+Q_1(x)\mI,\\
 Q_3(x)&=\alpha_1+\alpha_3x+\alpha_6x^{2}+\beta_3x^{3},\\
 Q_2(x)&=(\alpha_2+\alpha_3)+(\alpha_5+\alpha_6)x+(\beta_1+\beta_2+2\beta_3)x^{2},\\
 Q_1(x)&=\alpha_4+\beta_2x,
\end{aligned}
\end{align}
where $\mI$ is the identity operator: $\mI f(x)=f(x)$.
\begin{prop}
The generic Heun operator \eqref{eq:gen_contHeun} can be obtained as the most general
quadratic combination in the S-Heun generators \eqref{eq:barLM1M2R1R2} that does not
raise the degree of polynomials by more than one.
\end{prop}
Calling upon the reordering relations of Appendix \ref{sec:appendix}, it is seen that
the Heun operator generically factorizes as the product of a general S-Heun operator with
a stabilizing S-Heun operator:
\begin{align}\label{}
 \bar{W}=(\xi_1\bar{L}+\xi_2\bar{M}_1+\xi_3\bar{M}_2)
         (\eta_1\bar{L}+\eta_2\bar{M}_1+\eta_3\bar{M}_2+\eta_4\bar{R}_1+\eta_5\bar{R}_2)
        +\kappa.
\end{align}

\section{S-Heun operators on the linear grid}\label{sec:sheun}

We now come to one of the main topics of the paper, namely the S-Heun operators
defined on the linear grid.

\subsection{The stabilizing subset}\label{}

The subset of S-Heun operators that stabilizes the polynomials of a given degree is
$\{L,M_1,M_2\}$.  The most general quadratic combination of these operators can always be
reduced to an expression of the form
\begin{align}\label{}
 Q=\alpha_1{L}^{2}+\alpha_2{L}{M}_1+\alpha_3{L}{M}_2
        +\alpha_4{{M}_1}^{2}+\alpha_5{M}_1{M}_2+\alpha_6{{M}_2}^{2}
\end{align}
using the relations of Appendix \ref{sec:appendix}.  Substituting the expressions
\eqref{eq:LM1M2R1R2_def}, one sees that $Q$ is a second-order difference operator.  By
straightforward manipulations, the eigenvalue equation for $Q$ can be transformed into the
difference equation of the Continuous Hahn polynomials \cite{KoekoekLeskyetal2010}
\begin{align}\label{}
\begin{aligned}
 \mD P_n(\tilde{x};a,b,c,d)&=n(n+a+b+c+d-1)P_n(\tilde{x};a,b,c,d),\\
 \mD&=B(\tilde{x})T_+^{2}-[B(\tilde{x})+D(\tilde{x})]\mI+D(\tilde{x})T_-^{2},\\
 B(x)&=(c-ix)(d-ix),\qquad D(x)=(a+ix)(b+ix),
\end{aligned}
\end{align}
with $\tilde{x}=i\tfrac{x}{2}$ and where $a$, $b$, $c$, $d$ are given in terms of the
$\alpha_i$.
From this, we recognize that the key family of OPs related to these S-Heun operators is
the Continuous Hahn family.

\subsection{Continuous Hahn polynomials and their structure relations}\label{}

The following combinations of S-Heun operators
\begin{subequations}\label{eq:ttsmms_def}
\begin{align}\label{}
 \tau&=2L,\\
 \label{eq:taus_def}
 \tau^{(a,b,c,d)^{*}}&= \mu_1L+\mu_2M_1+\mu_3M_2+\mu_4R_1+\mu_5R_2,
\end{align}
with
\begin{align}
\begin{aligned}
 \mu_1&=\tfrac{1}{2}(1-(a+b+c+d))+(ab+cd),\\
 \mu_2&=\tfrac{1}{2}(a+b-c-d)-(ab-cd),\\
 \mu_3&=\tfrac{1}{2}(c+d-a-b),\\
 \mu_4&=-\tfrac{1}{4},\\
 \mu_5&=\tfrac{1}{2}(a+b+c+d)-\tfrac{3}{4}
\end{aligned}
\end{align}
turn out to be the forward and backward operators, while
\begin{align}\label{eq:mumus}
 \mu^{(a,b,c,d)}
    &=(d-a)L+(a+d-1)M_1+M_2,\\
 \mu^{(a,b,c,d)^{*}}
    &=(c-b)L+(b+c-1)M_1+M_2,
\end{align}
\end{subequations}
will act on polynomials as the contiguity relations. Indeed, these operators have the
following actions on the Continuous Hahn polynomials:
\begin{subequations}\label{eq:ttsmms_act}
\begin{align}
 \tau~P_n\left(i\tfrac{x}{2},a,b,c,d\right)
 &=i(n+a+b+c+d-1)P_{n-1}\left(i\tfrac{x}{2},
 a+\tfrac{1}{2},b+\tfrac{1}{2},c+\tfrac{1}{2},d+\tfrac{1}{2}\right),\\
 \tau^{(a,b,c,d)^{*}}~P_n\left(i\tfrac{x}{2},a,b,c,d\right)
 &=-i(n+1)P_{n+1}\left(i\tfrac{x}{2},
 a-\tfrac{1}{2},b-\tfrac{1}{2},c-\tfrac{1}{2},d-\tfrac{1}{2}\right),\\
 \mu^{(a,b,c,d)}~P_n\left(i\tfrac{x}{2},a,b,c,d\right)&=(n+a+d-1)
 P_{n}\left(i\tfrac{x}{2},a-\tfrac{1}{2},b+\tfrac{1}{2},c+\tfrac{1}{2},
 d-\tfrac{1}{2}\right),\\
 \mu^{(a,b,c,d)^{*}}~P_n\left(i\tfrac{x}{2},a,b,c,d\right)&=(n+b+c-1)
 P_{n}\left(i\tfrac{x}{2},a+\tfrac{1}{2},b-\tfrac{1}{2},c-\tfrac{1}{2},
 d+\tfrac{1}{2}\right).
\end{align}
\end{subequations}
The $4$ operators $\mu^{(a,b,c,d)}$, $\mu^{(a,b,c,d)^{*}}$, $\tau$, $\tau^{(a,b,c,d)^{*}}$
have been studied by Kalnins and Miller in \cite{KalninsMiller1987}.

We now indicate how the two bispectral operators are formed from the S-Heun operators. As
mentioned above, the Continuous Hahn difference operator can be formed by a quadratic
combination of the stabilizing generators. Moreover, we can provide
factorizations of this operator, either as a product of two contiguous operators or as the
product of the backward and forward operators:
\begin{align}\label{}
\begin{aligned}
 \mD&=\mu^{(a+\frac{1}{2},b-\frac{1}{2},c-\frac{1}{2},d+\frac{1}{2})}\mu^{(a,b,c,d)^{*}}
     -(a+d)(b+c-1)\\
    &=\mu^{(a-\frac{1}{2},b+\frac{1}{2},c+\frac{1}{2},d-\frac{1}{2})^{*}}\mu^{(a,b,c,d)}
     -(a+d-1)(b+c)\\
    &=\tau^{(a+\frac{1}{2},b+\frac{1}{2},c+\frac{1}{2},d+\frac{1}{2})^{*}}\tau\\
    &=\tau~\tau^{(a,b,c,d)^{*}}+2-(a+b+c+d).
\end{aligned}
\end{align}
The remaining bispectral operator $X$ is the multiplication by the variable $x$ in this
basis: $Xf(x)=xf(x)$. It appears as a quadratic combination in the S-Heun operators
\begin{align}\label{}
 X=[M_2,R_2].
\end{align}
The framework of S-Heun operators presented here is thus seen to unite the
symmetry techniques of Kalnins and Miller and the approach based on the bispectral
operators (see \cite{GenestVinetetal2014} for more general context).

\subsection{The Sklyanin-like algebra realized by the structure operators}\label{}

Let us now look at the algebraic relations obeyed by these operators.  On the one
hand, the pair of bispectral Continuous Hahn operators realizes the Hahn algebra
\cite{VinetZhedanov2019}.  On the other hand, the algebra formed by the $4$ linear
operators $\mu^{(a,b,c,d)}$, $\mu^{(a,b,c,d)^{*}}$, $\tau$, $\tau^{(a,b,c,d)^{*}}$ can be
seen as a degeneration of the Sklyanin algebra.

This algebra can be presented as follows. Write
$\nu=-\tfrac{1}{2}(a+b+c+d)$ and take
\begin{align}\label{eq:BMtdef}
\begin{aligned}
 A&=2(\nu+1)M_1-2M_2,\\
 B&=\tfrac{1}{2}(2\nu+1)(2\nu+3)L-R_1-(4\nu+3)R_2,\\
 C&=L,\\
 D&=M_1.
\end{aligned}
\end{align}
These are linear combinations of $\mu^{(a,b,c,d)}$, $\mu^{(a,b,c,d)^{*}}$, $\tau$,
$\tau^{(a,b,c,d)^{*}}$ that have been chosen in order to simplify the relations.
\begin{prop}
The elements $A$, $B$, $C$, $D$ obey the quadratic relations
\begin{subequations}\label{eq:Skl4}
\begin{gather}
 [C,D]=0,\qquad [A,C]=\{C,D\},\qquad
 [A,D]=\{C,C\},\label{eq:Skl4stab}\\[.5em]
 [B,C]=\{D,A\},\qquad
 [B,D]=\{C,A\},\qquad
 [B,A]=\{B,D\}.
\end{gather}
\end{subequations}
We shall refer to these relations as those of the $Skl_4$ algebra.
\end{prop}
The two quadratic Casimir elements are
\begin{align}\label{eq:Cas1Cas2}
 \Omega_1=D^{2}-C^{2},\qquad
 \Omega_2=A^{2}+D^{2}-\{B,C\}
\end{align}
and they take the following values in the realization:
\begin{align}\label{eq:valCas1Cas2}
 \Omega_1=1,\qquad
 \Omega_2=(2\nu+3)^{2}.
\end{align}
\begin{rmk}
The stabilizing subalgebra of $Skl_4$ \eqref{eq:Skl4stab}, which we shall denote by
$Skl_3$, has been identified in \cite{IyuduShkarin2018a} as the algebra
$T_7|_{(a,b)=(0,0)}$ whose relations are isomorphic to
\begin{align}\label{eq:T700}
 [x,y]=z^{2},\qquad [y,z]=0,\qquad [x,z]=zy.
\end{align}
It enjoys nice properties such as being Koszul, PBW, and being derived from a twisted
potential. That the above algebra is $Skl_3$ is seen by setting
$x=\tfrac{1}{2}A$, $y=D$, $z=C$.
\end{rmk}

We now explain that $Skl_4$ is a degeneration of the Sklyanin algebra.
We rewrite the $\tau^{(a,b,c,d)^{*}}$ in terms of $A$, $B$, $C$, $D$, using
$e_1=a+b+c+d$:
\begin{align}\label{eq:tauABCD}
 \tau^{(a,b,c,d)^{*}}=\tfrac{1}{4}(a+b-c-d)A+\tfrac{1}{4}B
 +\left[\tfrac{1}{8}(1-e_1)(1+e_1)+ab+cd\right]C
 +\left[\tfrac{1}{4}e_1(a+b-c-d)-ab+cd\right]D.
\end{align}
Two analogs of an identity due to Rains \cite{Rains2006} can be obtained for
$\tau^{(a,b,c,d)^{*}}$. These are the quasi-commutation relations:
\begin{align}
\label{eq:taustaustaustaus1}
 \tau^{(a+e,b,c,d-e)^{*}}\tau^{(a-\frac{1}{2},b+\frac{1}{2},c+\frac{1}{2},
 d-\frac{1}{2})^{*}}
 &=\tau^{(a,b,c,d)^{*}}\tau^{(a-\frac{1}{2}+e,b+\frac{1}{2},c+\frac{1}{2},
 d-\frac{1}{2}-e)^{*}},\\
\label{eq:taustaustaustaus2}
 \tau^{(a,b+e,c-e,d)^{*}}\tau^{(a+\frac{1}{2},b-\frac{1}{2},c-\frac{1}{2},
 d+\frac{1}{2})^{*}}
 &=\tau^{(a,b,c,d)^{*}}\tau^{(a+\frac{1}{2},b-\frac{1}{2}+e,c-\frac{1}{2}-e,
 d+\frac{1}{2})^{*}}.
\end{align}
\begin{prop}
Either of the quasi-commutation relation \eqref{eq:taustaustaustaus1},
\eqref{eq:taustaustaustaus2} repackages the relations \eqref{eq:Skl4} of the $Skl_4$
algebra.
\end{prop}
\textit{Proof:} Substituting the relation \eqref{eq:tauABCD} into
\eqref{eq:taustaustaustaus1} and bringing all terms to the rhs, one obtains ($u=b-c$,
$v=a-b-c+d$):
\begin{align}\label{eq:derivationSkl4p1}
\begin{aligned}
 0&=\tfrac{e}{4}\Big\{ \tfrac{1}{2}(AB-BA) + u(CB-BC)
 + \tfrac{1}{2}\left[(2-v)BD+vDB\right]
 + u\left[(2-v)AD+vDA-2(1-v)C^{2}\right]\\
 &~~-\tfrac{1}{4}\left[(v^{2}+4u^{2}-4v+3)AC-(v^{2}+4u^{2}-1)CA\right],\\
 &~~+\tfrac{1}{4}\left[v^{3}-4u^{2}v+8u^{2}-2v^{2}-v+2\right]CD
 -\tfrac{1}{4}\left[v^{3}-4u^{2}v-4v^{2}+3v\right]DC\Big\}.
\end{aligned}
\end{align}
The dependence on the free parameter $e$ factors out. Taking $v\to\infty$, one obtains
immediately that
\begin{align}\label{}
 CD-DC=0.
\end{align}
Also, taking $u\to0$ and $v\to0$, one gets
\begin{align}\label{}
 AB-BA=-2BD+\tfrac{3}{2}AC+\tfrac{1}{2}CA-CD.
\end{align}
Substituting these relations back in \eqref{eq:derivationSkl4p1} leads to
\begin{align}\label{eq:derivationSkl4p2}
\begin{aligned}
 0&=\tfrac{e}{4}\Big\{ u(CB-BC)
 + \tfrac{v}{2}\left[DB-BD\right]
 + u\left[(2-v)AD+vDA-2(1-v)C^{2}\right]\\
 &~~-\tfrac{1}{4}\left[(v^{2}+4u^{2}-4v)AC-(v^{2}+4u^{2})CA\right]
 + \tfrac{1}{4}\left[8u^{2}+2v^{2}-4v\right]CD\Big\}.
\end{aligned}
\end{align}
Repeating a similar process, the remaining relations of \eqref{eq:Skl4} are found.  A
similar derivation starting from \eqref{eq:taustaustaustaus2} instead yields the same
relations. \hfill$\square$

\subsection{Finite-dimensional representations}

It is known that finite-dimensional representations of the Hahn algebra relate to the Hahn
polynomials \cite{GranovskiiLutzenkoetal1992}. We now wish to obtain finite-dimensional
representations of the $Skl_4$ algebra; looking at \eqref{eq:BMtdef}, it is seen that one
needs $\nu$ to be either an integer or half-integer. It will be shown that this
corresponds in fact to a truncation of the Jacobi matrix of the Continuous Hahn
polynomials.

Let us write the condition ($\nu$ is either an integer or half-integer) as
\begin{align}\label{eq:truncation_lin}
 1-(a+b+c+d)=N,
\end{align}
where $N$ is a positive integer that corresponds to the maximal degree of the truncated
family of polynomials.

This truncation condition is known \cite{LemayVinetetal2016} to be the one that takes the
Wilson polynomials to the para-Racah polynomials.  In the present case, we start from the
Continuous Hahn OPs so the result of the truncation leads to a different family of
para-polynomials.
\begin{prop}
The polynomials that arise from the truncation condition \eqref{eq:truncation_lin} form a
basis that supports \penalty-10000$(N+1)$-dimensional representations of the degenerate
Sklyanin algebra $Skl_4$ and are identified as the para-Krawtchouk polynomials
\cite{VinetZhedanov2012}.
\end{prop}
We indicate below how the recurrence relation of the para-Krawtchouk polynomials is
obtained from that of the Continuous Hahn polynomials by imposing
\eqref{eq:truncation_lin}.

\subsubsection{$N=2j+1$ odd}\label{}

In the case where $N=2j+1$ is odd ($j$ is a non-negative integer), we parametrize the
truncation condition as follows
\begin{align}\label{}
 c=-a-j+e_1t,\qquad b=-d-j+e_2t
\end{align}
and then take the limit $t\to0$.
We shall choose $e_1=e_2$: this will lead to simpler expressions. The more general
solutions corresponding to $e_1\neq e_2$ can be recovered from the simpler solutions by
the procedure of isospectral deformations, see for instance \cite{LemayVinetetal2016a}.
Using the chosen parametrization, the recurrence coefficients $A_n$, $C_n$ appearing in
the recurrence relation of the Continuous Hahn polynomials
\begin{align}\label{}
\begin{aligned}
 (a+ix)P_n(x;a,b,c,d)&=A_nP_{n+1}(x;a,b,c,d)+C_nP_{n-1}(x;a,b,c,d)
 -(A_n+C_n)P_{n}(x;a,b,c,d),\\
 P_n(x;a,b,c,d)&=\frac{n!}{i^{n}(a+c)_n(a+d)_n}p_n(x;a,b,c,d)
\end{aligned}
\end{align}
become in the limit $t\to0$:
\begin{subequations}\label{eq:limitodd1}
\begin{align}
 A_n&=-\frac{(n-N)(n+a+d)}{2(2n-N)},\\
 C_n&=+\frac{n(n-N-a-d)}{2(2n-N)}.
\end{align}
\end{subequations}
Now take $\gamma$ to be
\begin{align}\label{}
 \gamma=(b+c)-(a+d),
\end{align}
it follows that \eqref{eq:limitodd1} can be rewritten in view of \eqref{eq:truncation_lin}
as
\begin{subequations}\label{eq:limitodd2}
\begin{align}
 A_n&=-\frac{1}{2}\frac{(N-n)(N-1-2n+\gamma)}{2(2n-N)},\\
 C_n&=-\frac{1}{2}\frac{n(N+1-2n-\gamma)}{2(2n-N)}.
\end{align}
\end{subequations}
These are recognized as the recurrence coefficients of the para-Krawtchouk polynomials in
the variable $-\tfrac{x}{2}$ introduced in \cite{VinetZhedanov2012}. These polynomials are
defined on the union of two linear lattices and the parameter $\gamma$ describes the
displacement of one lattice with respect to the other.

\subsubsection{$N=2j$ even}\label{}

In the case where $N=2j$ is even, we use the parametrization
\begin{align}\label{}
 c=-a-j+e_1t,\qquad b=-d-j+e_1t+1
\end{align}
and then take the limit $t\to0$. The recurrence coefficients in the recurrence relation of
the Continuous Hahn polynomials become
\begin{subequations}
\begin{align}
 A_n&=-\frac{(n-N)(n+a+d)}{2(2n-N+1)},\\
 C_n&=+\frac{n(n-N-a-d)}{2(2n-N-1)},
\end{align}
\end{subequations}
and upon writing
\begin{align}\label{}
 \gamma=1+(b+c)-(a+d),
\end{align}
we obtain
\begin{subequations}\label{eq:limitodd22}
\begin{align}
 A_n&=-\frac{1}{2}\frac{(N-n)(N-2-2n+\gamma)}{2(2n-N+1)},\\
 C_n&=-\frac{1}{2}\frac{n(N+2-2n-\gamma)}{2(2n-N-1)}.
\end{align}
\end{subequations}
These are the recurrence coefficients of the para-Krawtchouk polynomials in the
variable $-\tfrac{x}{2}$.  The expressions for the monic polynomials are given in
\cite{LemayVinetetal2016}.

\subsubsection{A remark on the truncation condition}\label{}

It can be checked that in the realization \eqref{eq:BMtdef}, applying the truncation
condition \eqref{eq:truncation_lin} seems to suggest that the raising operator $B$
annihilates the monomial $x^{N+1}$ and not $x^{N}$.
A priori, this means that the truncation condition amounts to
looking at ($N+2$)-dimensional representations of the algebra $Skl_4$, which would seem to
contradict the fact that the para-Krawtchouk polynomials were truncated to have degrees
at most $N$ (and thus to span a space of dimension $N+1$).

Looking at the situation more closely, one observes that $B$ indeed maps
para-Krawtchouk polynomial of maximal degree $N$ to a certain polynomial of degree $N+1$.
But this polynomial of degree $N+1$ corresponds to the characteristic polynomial of the
(upper block of the) truncated Jacobi matrix, hence it is null on the orthogonality grid
points. Keeping in mind that the para-Krawtchouk polynomials are the basis vectors for
the finite-dimensional representation of $Skl_4$, this characteristic polynomial thus
corresponds to a null vector. Therefore the dimension of the space on which the
representation of the $Skl_4$ algebra acts is indeed $N+1$.

\subsection{Recovering the associated Heun operator}\label{}

The Heun operator associated to the Continuous Hahn polynomials was implicitly
defined in \cite{VinetZhedanov2019}. This operator $W_{CH}$ is the most general second
order operator that acts on the discrete linear grid and maps polynomials of degree $n$
into polynomials of degree $n+1$. It can be expressed as
\begin{align}\label{eq:WCH_def}
 W_{CH}=\mA_1T_++\mA_0\mI+\mA_2T_-,
\end{align}
where $\mA_{1,2}$ are general polynomials of degree $3$ with the same leading order
coefficient, and $\mA_0+\mA_1+\mA_2=\pi_1(x)$, with $\pi_1(x)$ a general polynomial of
degree $1$.

We now consider the most general quadratic combination of S-Heun operators that does not
raise the degree of polynomials by more than one. Upon using the quadratic homogeneous
relations of Appendix \ref{sec:appendix}, this general combination can be brought into the
form
\begin{align}\label{}
 W=\alpha_1{L}^{2}+\alpha_2{L}{M}_1+\alpha_3{L}{M}_2
        +\alpha_4{{M}_1}^{2}+\alpha_5{M}_1{M}_2+\alpha_6{{M}_2}^{2}
        +\beta_1{M}_1{R}_2+\beta_2{M}_2{R}_1+\beta_3{M}_2{R}_2.
\end{align}
Substituting the expressions of the S-Heun basis operators \eqref{eq:LM1M2R1R2_def}, we
obtain
\begin{align}\label{}
\begin{aligned}
 &\qquad{W}=\mA_1T_+^{2}+\mA_0\mI+\mA_2T_-^{2},\\
 &\mA_1=\tfrac{1}{4}[-2\beta_2x^{3}+(\alpha_6-3\beta_2+\beta_3)x^{2}
  +(\alpha_3+\alpha_5+\alpha_6+\beta_1-\beta_2+\beta_3)x
  +(\alpha_1+\alpha_2+\alpha_3+\alpha_4+\alpha_5+\beta_1)],\\
 &\mA_2=\tfrac{1}{4}[-2\beta_2x^{3}+(\alpha_6+3\beta_2-\beta_3)x^{2}
  +(\alpha_3-\alpha_5-\alpha_6+\beta_1-\beta_2+\beta_3)x
  +(\alpha_1-\alpha_2-\alpha_3+\alpha_4+\alpha_5-\beta_1)],\\
 &\mA_0=(\beta_1+\beta_2+\beta_3)x+\alpha_4-(\mA_1+\mA_2).
\end{aligned}
\end{align}
\begin{prop}
The generic Heun-Continuous Hahn operator \eqref{eq:WCH_def} can be obtained as the most
general quadratic combination in the S-Heun generators \eqref{eq:LM1M2R1R2_def} that does
not raise the degree of polynomials by more than one.
\end{prop}
Using the relations of Appendix \ref{sec:appendix}, one can see that the
Heun operator generically factorizes as the product of a general S-Heun operator with a
stabilizing S-Heun operator:
\begin{align}\label{}
 {W}=(\xi_1{L}+\xi_2{M}_1+\xi_3{M}_2)
         (\eta_1{L}+\eta_2{M}_1+\eta_3{M}_2+\eta_4{R}_1+\eta_5{R}_2)
        +\kappa.
\end{align}

\section{The case of the $q$-linear grid}\label{sec:qlin}

We consider now the S-Heun operators associated to the $q$-linear (or exponential) grid.

\subsection{The stabilizing subspace}\label{}

The stabilizing subset of S-Heun operators is $\{\hat{L}, \hat{M}_1, \hat{M}_2\}$.  Using
the relations of Appendix \ref{sec:appendix}, it is always possible to reduce the most
general quadratic combination of these operators to
\begin{align}\label{}
 \hat{Q}=\alpha_1{\hat{L}}^{2}+\alpha_2{\hat{L}}{\hat{M}}_1+\alpha_3{\hat{L}}{\hat{M}}_2
  +\alpha_4{\hat{M}_1}{\!}^{2}+\alpha_5{\hat{M}}_1{\hat{M}}_2+\alpha_6{\hat{M}_2}{\!}^{2}.
\end{align}
Substituting the expressions \eqref{eq:hatLM1M2R1R2_def}, one recognizes $\hat{Q}$ as a
second-order $q$-difference operator whose eigenvalue problem can be cast as the
difference equation
\begin{align}\label{}
\begin{aligned}
 \hat{\mD}P_n(z;\alpha,\beta,\gamma;\tilde{q})
 &=(\tilde{q}^{-n}-1)(1-\alpha\beta\tilde{q}^{n+1})P_n(z;\alpha,\beta,\gamma;\tilde{q}),\\
 \hat{\mD}&=B(z)\hat{T}_+^{2}-[B(z)+D(z)]\mI+D(z)\hat{T}_-^{2},\\
 B(z)&=\frac{\alpha\tilde{q}(z-1)(\beta z-\gamma)}{z^{2}},\qquad
 D(z)=\frac{(z-\alpha\tilde{q})(z-\gamma\tilde{q})}{z^{2}}
\end{aligned}
\end{align}
of the Big $q$-Jacobi polynomials \cite{KoekoekLeskyetal2010} in base $\tilde{q}=q^{2}$,
making those the OPs associated to S-Heun operators on the exponential lattice.
\RED{
We note that there is a duality between the Continuous Dual $q$-Hahn and the Big
$q$-Jacobi polynomials \cite{KoornwinderMazzocco2018} that can be pictured as follows:
exchanging the degree with the variable in some way takes one family of polynomials
into the other (with transformed parameters). Thus, if we were to write the S-Heun
operators \eqref{eq:hatLM1M2R1R2_def} by replacing the variable with the degree in the
appropriate way, the Continuous Dual $q$-Hahn polynomials would arise instead.
}

\subsection{Big $q$-Jacobi polynomials and their structure relations}\label{}

Focusing on the structure and contiguity relations of the Big $q$-Jacobi polynomials,
we shall show how the set of S-Heun operators spans a space that contains the relevant
operators.  Let
\begin{subequations}\label{eq:hat_ttsmms_def}
\begin{align}\label{}
 \hat{\tau}&=(q-q^{-1})\hat{L},\\
 \label{eq:taus_defh}
 \hat{\tau}^{(a,b,c,d)^{*}}&
 =\mu_1\hat{L}+\mu_2\hat{M}_1+\mu_3\hat{M}_2+\mu_4\hat{R}_1+\mu_5\hat{R}_2,
\end{align}
with
\begin{align}
\begin{aligned}
 \mu_1&=-(q-q^{-1}),\\
 \mu_2&=(a+b)q^{-1}-q(c^{-1}+d^{-1}),\\
 \mu_3&=(a+b)-(c^{-1}+d^{-1}),\\
 \mu_4&=-abq^{-2}+q^{2}c^{-1}d^{-1},\\
 \mu_5&=-abq^{-1}+qc^{-1}d^{-1},
\end{aligned}
\end{align}
and
\begin{align}\label{eq:mumush}
 \hat{\mu}^{(a,b,c,d)}
    &=(q-q^{-1})L-(aq^{-1}-qd^{-1})M_1-(a-d^{-1})M_2,\\
 \hat{\mu}^{(a,b,c,d)^{*}}
    &=(q-q^{-1})L-(bq^{-1}-qc^{-1})M_1-(b-c^{-1})M_2.
\end{align}
\end{subequations}
The actions of these operators on the Big $q$-Jacobi polynomials
$P_n(z;\alpha,\beta,\gamma;q^{2})$ is best presented as follows. Let
\begin{align}\label{eq:hat_PhiPn}
 \Phi_n^{(a,b,c,d)}(z;\tilde{q})
 =P_n(az;ac\tilde{q}^{-1},bd\tilde{q}^{-1},ad\tilde{q}^{-1};\tilde{q}).
\end{align}
It is clear that the parameter $a$ is redundant. One has
$\Phi_n^{(1,\beta/\gamma,\alpha\tilde{q},\gamma\tilde{q})}(z;\tilde{q})
=P_n(z;\alpha,\beta,\gamma;\tilde{q})$. It is seen that
\begin{subequations}\label{eq:ttsmms_acth}
\begin{align}
 \hat{\tau}~\Phi_n^{(a,b,c,d)}(z;\tilde{q})
 &=\frac{aq(1-q^{-2n})(1-abcdq^{2n-2})}{(1-ad)(1-ac)}
 \Phi_{n-1}^{(aq,bq,cq,dq)}(z;\tilde{q}),\\
 \hat{\tau}^{(a,b,c,d)^{*}}~\Phi_n^{(a,b,c,d)}(z;\tilde{q})
 &=\frac{(ac-q^{2})(ad-q^{2})}{acdq}
 \Phi_{n+1}^{(aq^{-1},bq^{-1},cq^{-1},dq^{-1})}(z;\tilde{q}),\\
 \hat{\mu}^{(a,b,c,d)}~ \Phi_n^{(a,b,c,d)}(z;\tilde{q})
 &=\frac{q}{d}(1-adq^{-2})\Phi_n^{(aq^{-1},bq,cq,dq^{-1})}(z;\tilde{q}),\\
 \hat{\mu}^{(a,b,c,d)^{*}}~ \Phi_n^{(a,b,c,d)}(z;\tilde{q})
 &=-\frac{q(ad-q^{-2n})(1-bcq^{2n-2})}{c(1-ad)}
 \Phi_n^{(aq,bq^{-1},cq^{-1},dq)}(z;\tilde{q}).
\end{align}
\end{subequations}
The $4$ operators $\hat{\mu}^{(a,b,c,d)}$, $\hat{\mu}^{(a,b,c,d)^{*}}$, $\hat{\tau}$,
$\hat{\tau}^{(a,b,c,d)^{*}}$ built from linear combinations of S-Heun operators have been
studied by Kalnins and Miller in \cite{KalninsMiller1987}.

Let us further indicate how the bispectral operators show up in this context. As mentioned
above, the Big $q$-Jacobi difference operator appears as a quadratic combination of the
stabilizing generators. Moreover, one can actually provide factorizations of this operator
in terms of contiguity operators as well as backward and forward operators:
\begin{align}\label{}
\begin{aligned}
 \hat{\mD}&=\alpha\gamma q^{3}\mu^{(q,\frac{\beta}{\gamma q},\alpha q,\gamma q^{3})}
 \mu^{(1,\frac{\beta}{\gamma},\alpha q^{2},\gamma q^{2})^{*}}
   -(1-\gamma q^{2})(1-\tfrac{\alpha\beta}{\gamma})\\
  &=\alpha\gamma q^{3}\mu^{(q^{-1},\frac{\beta q}{\gamma},\alpha q^{3},\gamma q)^{*}}
 \mu^{(1,\frac{\beta}{\gamma},\alpha q^{2},\gamma q^{2})}
   -(1-\gamma)(1-\tfrac{\alpha\beta q^{2}}{\gamma})\\
  &=-\alpha\gamma q^{3}\hat{\tau}^{(q,\frac{\beta q}{\gamma},\alpha q^{3},\gamma
   q^{3})^{*}} \tau\\
  &=-\alpha\gamma q^{3}\hat{\tau}~\hat{\tau}^{(1,\frac{\beta}{\gamma},\alpha q^{2},\beta
   q^{2})^{*}}-(1-q^{2})(1-\alpha\beta).
\end{aligned}
\end{align}
The second bispectral operator $\hat{X}$ is the multiplication by the variable $z$:
$\hat{X}f(z)=zf(z)$. It also appears as the quadratic combination of S-Heun operators:
\begin{align}\label{}
 \hat{X}=\hat{M}_2\hat{R_1}-\hat{M}_1\hat{R}_2.
\end{align}
The S-Heun operators thus underscore much of the characterization of the Big $q$-Jacobi
operators.

\subsection{The Sklyanin-type algebra realized by the structure operators}\label{}

The pair of bispectral Big $q$-Jacobi operators is known to realize the Big $q$-Jacobi
algebra \cite{TsujimotoVinetetal2017, BaseilhacVinetetal2020}.  The algebra generated by
the $4$ linear operators $\hat{\mu}^{(a,b,c,d)}$, $\hat{\mu}^{(a,b,c,d)^{*}}$,
$\hat{\tau}$, $\hat{\tau}^{(a,b,c,d)^{*}}$ is a familiar degeneration of the Sklyanin
algebra \cite{Sklyanin1983}.

Denote
$q^{-\nu}=(abcd)^{\frac{1}{4}}$ and form
\begin{align}\label{eq:uqsl2_real}
\begin{aligned}{}
 \hat{A}&=q^{-\nu}(\hat{M}_1+q\hat{M_2}),\\
 \hat{B}&=\frac{1}{2(q-q^{-1})}[q^{2\nu}(\hat{R_1}+q^{-1}\hat{R_2})
  -q^{-2\nu}(\hat{R_1}+q\hat{R_2})],\\
 \hat{C}&=2\hat{L},\\
 \hat{D}&=q^{\nu}(\hat{M}_1+q^{-1}\hat{M_2}).
\end{aligned}
\end{align}
\begin{prop}
The operators $\hat{A}$, $\hat{B}$, $\hat{C}$, $\hat{D}$ obey the quadratic relations
\begin{subequations}\label{eq:uqsl2}
\begin{align}\label{eq:homuqsl2}
\begin{aligned}{}
 \hat{A}\hat{B}=q\hat{B}\hat{A}, \quad & \quad
 \hat{B}\hat{D}=q\hat{D}\hat{B}, \qquad
 \hat{C}\hat{A}=q\hat{A}\hat{C,} \qquad
 \hat{D}\hat{C}=q\hat{C}\hat{D},\\
 &[\hat{B},\hat{C}] = \frac{\hat{A}^2-\hat{D}^2}{q-q^{-1}}, \qquad
 [\hat{A},\hat{D}]=0
\end{aligned}
\end{align}
\RED{
along with the additional relation
\begin{align}\label{}
 \hat{A}\hat{D}=\hat{D}\hat{A}=1
\end{align}
}
\end{subequations}
which define $U_q(\sl_2)$.
\end{prop}
When $\nu$ is an integer or a half-integer, one obtains finite-dimensional representations
of $U_q(\sl_2)$ of dimension $2\nu+1$. In that case, the maximal degree of the polynomials
obtained from the action of the raising operator $\hat{B}$ is $N$.
\begin{rmk}
The $q\to1$ limit of this realization yields the $\sl_2$ commutation
relations. In fact \eqref{eq:uqsl2_real} tends to the differential Bargmann
realization of $\sl_2$. Under the limit, the $q$-linear grid becomes the continuum, and
the above combinations of shift operators turn into differential operators.
\end{rmk}
\RED{
\begin{rmk}
The algebra \eqref{eq:barSkl4} has been obtained in \cite{LeBruynSmith1993} as a so-called
``homogenized $\sl_2$ algebra'' $H(\sl_2)$. Many algebras of a similar type with $4$
generators $A$, $B$, $C$, $D$, and $D$ central, have been studied in
\cite{LeBruynSmithetal1996}.  A quantization of $H(\sl_2)$ which is isomorphic to the
algebra with relations \eqref{eq:homuqsl2} and which can be seen as a homogenization of
$U_q(\sl_2)$ has been studied in \cite{ChirvasituSmithetal2018}.
\end{rmk}
}

\subsection{Finite-dimensional representations}

We now wish to obtain finite-dimensional representations of $U_q(\sl_2)$ corresponding to
a particular truncation of the Jacobi matrix of the Big $q$-Jacobi polynomials. As
mentioned previously, this can be accomplished by taking $\nu$ to be either an integer or
a half-integer.  In order to do so, we are led to take \cite{LemayVinetetal2018}
\begin{align}\label{eq:truncation_hat}
 \sqrt{abcd}=q^{1-N},
\end{align}
where $N$ is a positive integer that corresponds to the maximal degree of the truncated
family of polynomials.
\begin{prop}
The polynomials that arise from the truncation condition \eqref{eq:truncation_hat} form a
basis that supports \penalty-10000$(N+1)$-dimensional representations of $U_q(\sl_2)$ in
the realization \eqref{eq:uqsl2_real}.  The $q$-para-Krawtchouk polynomials
\cite{TsujimotoVinetetal2017} are the ones that arise from this truncation condition.
\end{prop}
We show below how their recurrence relation is obtained from the one of the Big $q$-Jacobi
polynomials.

\subsubsection{$N=2j+1$ odd}\label{}

In the case where $N=2j+1$ is odd, we write
\begin{align}\label{eq:qtruncodd}
 d=a^{-1}q^{-2j+e_1t},\qquad b=c^{-1}q^{-2j+e_1t}
\end{align}
and then take the limit $t\to0$.
Using this parametrization, the recurrence relation of the Big $q$-Jacobi polynomials
\begin{align}\label{}
 zP_n(z;a,b,c;\tilde{q})&=A_nP_{n+1}(z;a,b,c;\tilde{q})+C_nP_{n-1}(z;a,b,c;\tilde{q})
 +[1-(A_n+C_n)]P_{n}(z;a,b,c;\tilde{q})
\end{align}
has for coefficients
\begin{subequations}\label{eq:hat_limitodd1}
\begin{align}
 A_n&=+\frac{(1-acq^{2n})(1-q^{2n-2N})}{(1+q^{2n-N+1})(1-q^{4n-2N})},\\
 C_n&=-\frac{q^{2n-N-1}(1-q^{2n})(ac-q^{2n-2N})}{(1+q^{2n-N-1})(1-q^{4n-2N})}
\end{align}
\end{subequations}
after the use of \eqref{eq:qtruncodd} and the limit $t\to0$.
Now letting
\begin{align}\label{}
 ac=c_3q^{2}
\end{align}
it follows that \eqref{eq:hat_limitodd1} can be rewritten as
\begin{subequations}\label{eq:hat_limitodd2}
\begin{align}
 A_n&=+\frac{(1-c_3q^{2n+2})(1-q^{2n-2N})}{(1+q^{2n-N+1})(1-q^{4n-2N})},\\
 C_n&=-\frac{q^{2n-N+1}(1-q^{2n})(c_3-q^{2n-2N-2})}{(1+q^{2n-N-1})(1-q^{4n-2N})},
\end{align}
\end{subequations}
and one recognizes the recurrence coefficients of the $q$-para-Krawtchouk polynomials in
the base $\tilde{q}=q^{2}$ introduced in \cite{TsujimotoVinetetal2017} when $N$ is odd.
These polynomials are defined on the union of two $q$-linear lattices and the parameter
$c_3$ describes the shift of one lattice with respect to the other.

\subsubsection{$N=2j$ even}\label{}

In the case where $N=2j$ is even, we take
\begin{align}\label{}
 d=a^{-1}q^{-2j+e_1t},\qquad b=c^{-1}q^{-2j+e_2t+2}
\end{align}
which ensures \eqref{eq:truncation_hat} in the limit $t\to0$.  Using this parametrization
and after letting $t\to0$, the recurrence coefficients of the Big $q$-Jacobi polynomials
become
\begin{subequations}\label{eq:hat_limiteven1}
\begin{align}
 A_n&=+\frac{(1-acq^{2n})(1-q^{2n-2N})}{(1+q^{2n-N})(1-q^{4n-2N+2})},\\
 C_n&=-\frac{q^{2n-N-2}(1-q^{2n})(ac-q^{2n-2N})}{(1+q^{2n-N})(1-q^{4n-2N-2})},
\end{align}
\end{subequations}
and upon letting
\begin{align}\label{}
 ac=c_3q^{2}
\end{align}
$A_n$ and $C_n$ can be rewritten as
\begin{subequations}\label{eq:hat_limiteven2}
\begin{align}
 A_n&=+\frac{(1-c_3q^{2n+2})(1-q^{2n-2N})}{(1+q^{2n-N})(1-q^{4n-2N+2})},\\
 C_n&=-\frac{q^{2n-N}(1-q^{2n})(c_3-q^{2n-2N-2})}{(1+q^{2n-N})(1-q^{4n-2N-2})}.
\end{align}
\end{subequations}
These are the recurrence coefficients of the $q$-para-Krawtchouk polynomials in the base
$\tilde{q}=q^{2}$ for $N$ even. For more detail, see \cite{TsujimotoVinetetal2017}.

\subsubsection{A remark on the truncation condition}\label{}

There is once again an apparent mismatch in the dimensions of the representations of the
algebra and those of the representation basis. The same remark as the one made in the
preceding section applies here.  It can be checked that in the realization
\eqref{eq:uqsl2_real}, applying the truncation condition \eqref{eq:truncation_hat} seems
to suggest that the raising operator $\hat{B}$ annihilates the monomial $z^{N+1}$ and not
$z^{N}$, which means that the truncation condition leads to representations of the algebra
$U_q(\sl_2)$ of dimension $N+2$. This would contradict the fact that the
$q$-para-Krawtchouk polynomials were truncated to a maximal degree $N$ (and thus span a
space of dimension $N+1$).

It can be observed that $\hat{B}$ maps the $q$-para-Krawtchouk polynomial of degree $N$ to
a polynomial of degree $N+1$. The resulting polynomial is the characteristic polynomial of
the (upper block of the) truncated Jacobi matrix, hence it is again null on the
orthogonality grid points.  In the representation basis with which we are working (i.e.
where the $q$-para-Krawtchouk polynomials are the basis elements), this characteristic
polynomial corresponds to a null vector. Hence, the dimension of the space on which the
realization of the $U_q(\sl_2)$ algebra acts is indeed $N+1$.

\subsection{Recovering the related Heun operator}\label{}

The Heun operator associated to the Big $q$-Jacobi polynomials is given in
\cite{BaseilhacVinetetal2020} and had also been introduced previously in
\cite{Takemura2018}. This operator $W_{BJ}$ is the most general second order
$q$-difference operator that acts on the $q$-linear grid and maps polynomials of degree
$n$  into polynomials of degree $n+1$. Its expression is
\begin{align}\label{eq:WBJ_def}
 W_{BJ}=\mA_1\hat{T}_++\mA_0\mI+\mA_2\hat{T}_-,
\end{align}
where
\begin{align}\label{}
 \mA_1=\frac{\pi_3(z)}{z^{2}},\qquad \mA_2=\frac{\tilde{q}\pi_3(z)+z\pi_2(z)}{z^{2}}
\end{align}
and $\mA_0+\mA_1+\mA_2=\pi_1(z)$, with $\pi_k(z)$ a generic polynomial of degree $k$ and
$\tilde{q}$ the base.

Let us consider the most general quadratic combination of S-Heun operators that does not
raise the degree of polynomials by more than one. Using the quadratic homogeneous
relations of Appendix \ref{sec:appendix}, we arrive at
\begin{align}\label{}
 W=\alpha_1\hat{L}^{2}+\alpha_2\hat{L}\hat{M}_1+\alpha_3\hat{L}\hat{M}_2
        +\alpha_4\hat{M}_1{\!}^{2}+\alpha_5\hat{M}_1\hat{M}_2+\alpha_6\hat{M}_2{\!}^{2}
        +\beta_1\hat{M}_1\hat{R}_2+\beta_2\hat{M}_2\hat{R}_1+\beta_3\hat{M}_2\hat{R}_2.
\end{align}
Substituting the expressions \eqref{eq:hatLM1M2R1R2_def} for the generators we obtain
\begin{align}\label{}
\begin{aligned}
 &\qquad{W}=\mA_1\hat{T}_+^{2}+\mA_0\mI+\mA_2\hat{T}_-^{2},\\
 &\mA_1=\frac{1}{z^{2}(1-q^{2})^{2}}[(q\alpha_1)+(q^{2}\alpha_3-q\alpha_2)z
 +(q^{2}\alpha_6-q\alpha_5+\alpha_4)z^{2}
 +(q^{3}\beta_3-q^{2}\beta_1-q^{2}\beta_2)z^{3}],\\
 &\mA_2=\frac{1}{z^{2}(1-q^{2})^{2}}[(q^{3}\alpha_1)+(q^{2}\alpha_3-q^{3}\alpha_2)z
 +(q^{2}\alpha_6-q^{3}\alpha_5+q^{4}\alpha_4)z^{2}
 +(q\beta_3-q^{2}\beta_1-q^{2}\beta_2)z^{3}],\\
 &\mA_0=\beta_2z+\alpha_4-(\mA_1+\mA_2).
\end{aligned}
\end{align}
\begin{prop}
The generic Heun-Big $q$-Jacobi operator \eqref{eq:WBJ_def} (with base $q^{2})$ can be
obtained as the most general quadratic combination in the S-Heun generators
\eqref{eq:hatLM1M2R1R2_def} that does not raise the degree of polynomials by more than
one.
\end{prop}
Moreover, using the relations of Appendix \ref{sec:appendix}, we see that the
Heun operator typically factorizes as the product of a raising S-Heun operator with a
stabilizing S-Heun operator:
\begin{align}\label{}
 \hat{W}=(\xi_1\hat{L}+\xi_2\hat{M}_1+\xi_3\hat{M}_2)
         (\eta_1\hat{L}+\eta_2\hat{M}_1+\eta_3\hat{M}_2+\eta_4\hat{R}_1+\eta_5\hat{R}_2)
        +\kappa.
\end{align}

\section{Connections between the different cases}\label{sec:connection}

It is well known that the three grids on which we have defined S-Heun operators can be
obtained as limiting cases or contractions of the Askey--Wilson grid. We now observe that
this translates into limits/contractions of the associated Sklyanin algebras.

Let us denote the points of the Askey--Wilson grid by
\begin{align}\label{}
 \lambda_s=z_{s}+z_{s}^{-1},\qquad z_{s}=q^{s}.
\end{align}
The associated Sklyanin algebra was introduced in \cite{GorskyZabrodin1993} as the
trigonometric degeneration of the Sklyanin algebra \cite{Sklyanin1983} and was studied
from the perspective of S-Heun operators in \cite{GaboriaudTsujimotoetal2020}. The
defining relations read
\begin{gather}\label{eq:eqp1}
\begin{gathered}
 \textbf{D}\textbf{C}=q\textbf{C}\textbf{D},\qquad
 \textbf{C}\textbf{A}=q\textbf{A}\textbf{C},\qquad
 [\textbf{A},\textbf{D}]=\frac{(q-q^{-1})^{3}}{4}\textbf{C}^{2},\\
 ~~~[\textbf{B},\textbf{C}]=\frac{\textbf{A}^{2}-\textbf{D}^{2}}{q-q^{-1}},\\
 \textbf{A}\textbf{B}-q\textbf{B}\textbf{A}=
  \,q\textbf{D}\textbf{B}-\textbf{B}\textbf{D}
  =-\frac{q^{2}-q^{-2}}{4}(\textbf{D}\textbf{C}-\textbf{C}\textbf{A}).
\end{gathered}
\end{gather}
The $q$-linear (or exponential) grid
\begin{align}\label{}
 \lambda_s=z_s,\qquad z_s=q^{s}
\end{align}
is obtained from the Askey--Wilson one in the asymptotic expansion $z_s\to\infty$ and the
same limit takes the Askey--Wilson polynomials into the Big $q$-Jacobi OPs. At the level
of the algebras, this corresponds to the following contraction. Writing
\begin{align}\label{eq:contract_alg}
 \textbf{A}=\epsilon\hat{A},\qquad
 \textbf{B}=\hat{B},\qquad
 \textbf{C}=\epsilon^{2}\hat{C},\qquad
 \textbf{D}=\epsilon\hat{D}
\end{align}
and taking $\epsilon\to0$, one recovers $U_q(\sl_2)$:
\begin{gather}\label{eq:eqp2}
\begin{gathered}{}
 \hat{A}\hat{B}=q\hat{B}\hat{A}, \qquad
 \hat{B}\hat{D}=q\hat{D}\hat{B}, \qquad
 \hat{C}\hat{A}=q\hat{A}\hat{C,} \qquad
 \hat{D}\hat{C}=q\hat{C}\hat{D},\\
 [\hat{B},\hat{C}] = \frac{\hat{A}^2-\hat{D}^2}{q-q^{-1}}, \qquad
 [\hat{A},\hat{D}]=0.
\end{gathered}
\end{gather}
We now compare the discrete linear grid to the continuum. A rescaling similar to the one
discussed above takes this grid to the real line. This also takes the Continuous
Hahn polynomials into the Jacobi ones. From the perspective of the algebras,
\eqref{eq:contract_alg} will relate one algebra to the other.  The Sklyanin algebra
\eqref{eq:Skl4} associated to the discrete grid is
\begin{gather}\label{eq:eqp3}
\begin{gathered}{}
 [{C},{D}]=0,\qquad [{A},{C}]=\{{C},{D}\},\qquad
 [{A},{D}]=\{{C},{C}\},\\
 [{B},{C}]=\{{D},{A}\},\qquad
 [{B},{D}]=\{{C},{A}\},\qquad
 [{B},{A}]=\{{B},{D}\}
\end{gathered}
\end{gather}
and upon writing
\begin{align}\label{}
 {A}=\epsilon\bar{A},\qquad
 {B}=\bar{B},\qquad
 {C}=\epsilon^{2}\bar{C},\qquad
 {D}=\epsilon\bar{D}
\end{align}
and taking $\epsilon\to0$, we recover
\begin{gather}\label{eq:eqp4}
\begin{gathered}{}
 [\bar{C},\bar{D}]=0,\qquad
 [\bar{A},\bar{C}]=-\bar{C}\bar{D},\qquad
 [\bar{A},\bar{D}]=0,\\
 [\bar{B},\bar{C}]=-2\bar{A}\bar{D},\qquad
 [\bar{A},\bar{B}]=\bar{B}\bar{D},\qquad
 [\bar{B},\bar{D}]=0.
\end{gathered}
\end{gather}
We recall that the latter algebra is essentially the $\sl_2$ Lie algebra with a central
element $D$.

We have so far discussed the following contractions, denoted by full arrows:
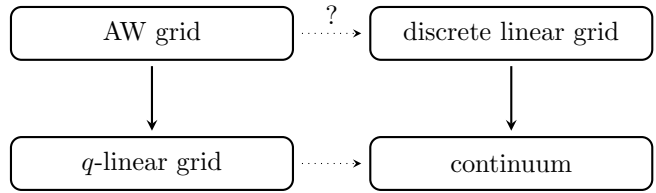
\tikzset{
    punkt/.style={
           rectangle,
           rounded corners,
           draw=black, thick,
           text width=10em,
           minimum height=2em,
           text centered},
}
\begin{align*}
\begin{aligned}
 \begin{tikzpicture}[node distance=1cm, auto,]
  \tikzstyle{arrow} = [thick,->,>=stealth,shorten <=2pt,shorten >=2pt]
  \tikzstyle{arrowd} = [dotted,->,>=stealth,shorten <=3pt,shorten >=3pt]
  \node[punkt] (law) {AW grid};
  \node[punkt,below=of law] (lql) {$q$-linear grid};
  \node[punkt,right=of law] (ldl) {discrete linear grid};
  \node[punkt,below=of ldl] (lcl) {continuum};
  \draw [arrowd] (law) -- node[anchor=south]{?} (ldl);
  \draw [arrow] (law) -- (lql);
  \draw [arrow] (ldl) -- (lcl);
  \draw [arrowd] (lql) -- (lcl);
 \end{tikzpicture}
\end{aligned}
\end{align*}
One could wonder if it is possible to complete the diagram with the dotted arrows.
The bottom arrow is easy to add: this amounts to taking the limit $q\to1$. This limit
takes the $q$-linear grid to the continuum, the Big $q$-Jacobi polynomials to the
Jacobi polynomials, and at the level of the algebra, it takes $U_q(\sl_2)$ to $\sl_2$.

The details corresponding to the upper arrow remain to be worked out. It is likely that an
intermediary step related to the quadratic grid $\lambda_s=s^{2}$ should be required.
Indeed, it is known that the $q\to1$ limit of the Askey--Wilson grid leads to the quadratic
grid. It should thus be possible to apply the S-Heun construction to the quadratic grid;
the related polynomials should be those of Wilson, and the related Sklyanin algebra would
stand in between the one of Askey--Wilson type \eqref{eq:eqp1} and the one of the discrete
linear type \eqref{eq:Skl4}.

\section{Conclusion}\label{sec:conclusion}

The results of this paper are summarized as follows.
We have introduced S-Heun operators on linear and $q$-linear grids. These operators are
special cases of second order Heun operators with no diagonal term.  On the real line and
the discrete and $q$-linear grids, the sets of five S-Heun operators were constructed and
shown to be related to the Jacobi, Continuous Hahn and Big $q$-Jacobi polynomials
respectively. These S-Heun operators were also shown to encompass the bispectral and
structure operators for each family of orthogonal polynomials. A presentation of the
relations for the four structure operators of Kalnins and Miller was given in each case
and identified as realizing degenerations, contractions or limits of the Sklyanin algebra.
For the discrete and $q$-linear grids, the finite-dimensional representations of the
Sklyanin-type algebras were obtained from a truncation condition on the Jacobi matrix of
the associated polynomials; this yielded the para-Krawtchouk and $q$-para-Krawtchouk
polynomials as bases of the finite representations and provided algebraic interpretations
of these sets of OPs that had so far been missing.

The Sklyanin-like algebra related to the discrete linear grid \eqref{eq:Skl4} has a simple
presentaton and a detailed study of its representation theory would be interesting.  It
would also be instructive to examine the types of Sklyanin algebra that the S-Heun
operators on the quadratic grid would lead to. We plan on undertaking this in the near
future. Note that we have restricted ourselves to Heun operators defined by actions on
polynomials. The exploration of the generalizations that result from the extension to
spaces of rational functions have been initiated in \cite{TsujimotoVinetetal2019} and
should be actively pursued in the S-Heun framework in particular.

\subsection*{Acknowledgments}\label{}

The authors would like to thank Jean-Michel Lemay for useful discussions
\RED{
as well as Paul Smith for enlightening correspondence}.
JG holds an Alexander-Graham-Bell scholarship from the Natural Sciences and
Engineering Research Council of Canada (NSERC).
The research of LV is funded in part by a Discovery Grant from NSERC.
AZ gratefully holds a CRM-Simons professorship and his work is supported by the National
Science Foundation of China (Grant No.11771015).

\subsection*{Data availability}\label{}

The data that support the findings of this study are available within the article.

\appendix
\section{The homogeneous quadratic algebraic relations}\label{sec:appendix}

The $14$ quadratic homogeneous relations associated to all three sets of $5$ S-Heun
operators are collected here. One notes that all three sets of relations display a similar
structure. These relations can be thought of as reordering relations and are especially
useful when considering the most general quadratic combinations in the generators.

\subsection{The continuum}\label{}

The relations between the S-Heun operators $\bar{L}$, $\bar{M}_1$, $\bar{M}_2$,
$\bar{R}_1$, $\bar{R}_2$ defined in \eqref{eq:barLM1M2R1R2} can be presented as the
fourteen following relations:
\begin{align}\label{}
\begin{aligned}{}
 \bar{M}_1\bar{L}&=\bar{L}\bar{M}_1,\\
 \bar{M}_2\bar{L}&=\bar{L}\bar{M}_2-\bar{M}_1\bar{L},\\
 \bar{M}_2\bar{M}_1&=\bar{M}_1\bar{M}_2,\\
 \bar{M}_1{\!}^{2}&=1,
\end{aligned}
\hspace{4em}
\begin{aligned}{}
 \bar{L}\bar{R}_1&=1+\bar{M}_1\bar{M}_2,\\
 \bar{L}\bar{R}_2&=\bar{M}_2{\!}^{2}+\bar{M}_1\bar{M}_2,\\
 \bar{R}_1\bar{L}&=\bar{M}_1\bar{M}_2,\\
 \bar{R}_2\bar{L}&=\bar{M}_2{\!}^{2}-\bar{M}_1\bar{M}_2,\\
 \bar{R}_2\bar{R}_1&=\bar{R}_1\bar{R}_2+\bar{R}_1{\!}^{2},
\end{aligned}
\hspace{4em}
\begin{aligned}{}
 \bar{R}_1\bar{M}_1&=\bar{M}_2\bar{R}_1-\bar{M}_1\bar{R}_2,\\
 \bar{R}_2\bar{M}_1&=\bar{M}_1\bar{R}_2,\\
 \bar{R}_1\bar{M}_2&=\bar{M}_1\bar{R}_2,\\
 \bar{R}_2\bar{M}_2&=\bar{M}_2\bar{R}_2-\bar{M}_1\bar{R}_2,\\
 \bar{M}_1\bar{R}_1&=\bar{M}_2\bar{R}_1-\bar{M}_1\bar{R}_2.
\end{aligned}
\end{align}

\subsection{The discrete linear grid}\label{}

Here are the relations between the S-Heun operators
$L$, $M_1$, $M_2$, $R_1$, $R_2$ that have been defined in \eqref{eq:LM1M2R1R2_def}:
\begin{align}
\begin{aligned}{}
 M_1L&=LM_1,\\
 M_2L&=LM_2-LM_1,\\
 M_2M_1&=M_1M_2-L^{2},\\
 {M_1}^{2}&=1+L^{2},
\end{aligned}
\hspace{3em}
\begin{aligned}{}
 LR_1&=1-2{M_2}^{2}-M_1M_2,\\
 LR_2&=1+M_1M_2,\\
 R_1L&=3M_1M_2-3L^{2}-2{M_2}^{2},\\
 R_2L&=M_1M_2-L^{2},\\
 R_2R_1&=2{R_2}^{2}+R_1R_2-4{M_2}^{2},
\end{aligned}
\hspace{3em}
\begin{aligned}{}
 R_1M_1&=3M_1R_2-2M_2R_2-3LM_1,\\
 R_1M_2&=2M_2R_2-3M_1R_2+3LM_2+M_2R_1,\\
 R_2M_1&=M_1R_2-LM_1,\\
 R_2M_2&=M_2R_2-M_1R_2+LM_2,\\
 M_1R_1&=3M_1R_2-2M_2R_2-4LM_2.
\end{aligned}
\end{align}

\subsection{The $q$-linear grid}\label{}

We remind the reader that the $q$-number $2$ is written as $[2]_q=q+q^{-1}$. The S-Heun
operators $\hat{L}$, $\hat{M}_1$, $\hat{M}_2$, $\hat{R}_1$, $\hat{R}_2$ defined in
\eqref{eq:hatLM1M2R1R2_def} obey the fourteen quadratic relations:
\begin{align}\label{}
\begin{aligned}{}
 \hat{M}_1\hat{L}&=[2]_q\hat{L}\hat{M}_1+\hat{L}\hat{M}_2,\\
 \hat{M}_2\hat{L}&=-\hat{L}\hat{M}_1,\\
 \hat{M}_2\hat{M}_1&=\hat{M}_1\hat{M}_2,\\
 [2]_q\hat{M}_1\hat{M}_2&=1-\hat{M}_1{\!}^{2}-\hat{M}_2{\!}^{2},
\end{aligned}
\hspace{3em}
\begin{aligned}{}
 \hat{L}\hat{R}_1&=1-\hat{M}_2{\!}^{2},\\
 \hat{L}\hat{R}_2&=[2]_q\hat{M}_2{\!}^{2}+\hat{M}_1\hat{M}_2,\\
 \hat{R}_1L&=1-\hat{M}_1{\!}^{2},\\
 \hat{R}_2\hat{L}&=-\hat{M}_1\hat{M}_2,\\
 [2]_q\hat{R}_1\hat{R}_2&=-\hat{R}_1{\!}^{2}-\hat{R}_2{\!}^{2},
 \end{aligned}
 \hspace{3em}
\begin{aligned}{}
 \hat{R}_1\hat{M}_1&=-[2]_q{\!}^{2}\hat{M}_1\hat{R}_2-[2]_q\hat{M_2}\hat{R}_2
  +\hat{M_2}\hat{R}_1,\\
 \hat{R}_1\hat{M}_2&=[2]_q\hat{M_1}\hat{R}_2+\hat{M_2}\hat{R}_2,\\
 \hat{R}_2\hat{M}_1&=[2]_q\hat{M_1}\hat{R}_2+\hat{M_2}\hat{R}_2,\\
 \hat{R}_2\hat{M}_2&=-\hat{M_1}\hat{R}_2,\\
 \hat{M_1}\hat{R}_1&=-[2]_q\hat{M_1}\hat{R}_2-\hat{M_2}\hat{R}_2.
\end{aligned}
\end{align}

\printbibliography

\end{document}